\documentclass[12pt]{article}

\usepackage{amssymb,amsmath,amsfonts,amsthm, amscd, mathrsfs,helvet}
\usepackage{bm}
\usepackage{graphicx,verbatim}
\usepackage{psfrag}
\usepackage[all]{xy}
\usepackage{layout}
\usepackage{wrapfig}
\usepackage{color}

\theoremstyle{plain}

\newcommand{\tensor}[1]{{\bf \underline{#1}}}

\DeclareMathOperator{\tr}{tr} \DeclareMathOperator{\str}{str}  

\def\1{{\bf \underline 1}}
\def\2{{\bf \underline 2}}

\def\A{{\cal A}}
\def\C{{\cal C}}
\def\D{{\cal D}}
\def\H{{\cal H}}
\def\K{{\cal K}}
\def\L{{\cal L}}
\def\P{{\cal P}}

\def\T{{\cal T}}

\def\L{\mathcal{L}}

\def\gd{\mathfrak{g}^{\ast}}

\def\ha{\mbox{\small $\frac{1}{2}$}}

\numberwithin{equation}{section}

\begin{document}

\begin{titlepage}
\begin{flushright}
{\bf October 2009} \\
IPhT-t09/143\\

\end{flushright}
\begin{centering}
\vspace{6mm} %
{\Large {\bf Hamiltonian dynamics and the hidden\\ symmetries of the $\bm{AdS_5 \times S^5}$ superstring}}\\

\vspace{6mm}
{\large Beno\^{\i}t Vicedo}\\
\vspace{4mm}
{\it Ecole Normale Sup\'erieure, LPT,\\
75231 Paris CEDEX-5, France \& \\
\vspace{2mm}
Institut de Physique Th\'eorique, C.E.A.-Saclay,\\
F-91191 Gif-sur-Yvette, France}\\
\vspace{2mm}
\small{\tt benoit.vicedo@cea.fr}

\vspace{15mm} 
{\bf Abstract} \\
\vspace{5mm} 

\end{centering}

We construct the Lax connection of the Green-Schwarz superstring in $AdS_5 \times S^5$ within the Hamiltonian formalism and obtain precisely that used in 0810.4136. It differs in a crucial way from the Bena-Polchinski-Roiban connection by terms proportional to the Hamiltonian constraints. These extra terms ensure firstly that the integrals of motion are all first class and secondly that the Lax connection is flat in the strong sense.

\end{titlepage}

\setcounter{page}{2}

\input{epsf}

The concept of integrability has so far played a vital role in deepening our understanding of both sides of the AdS/CFT correspondence. On the string theory side the first instance of classical integrability appeared in the seminal work of Bena, Polchinski and Roiban \cite{BPR}. There they constructed a Lax connection for the Green-Schwarz superstring on $AdS_5 \times S^5$ whose desirable features are its dependence on a \textit{spectral parameter} and its \textit{on-shell flatness}, meaning that it satisfies the zero curvature equation if and only if the fields of the theory satisfy their equations of motion. Such a property is perfect for solving the field equations since the zero curvature equation with spectral parameter carries a wealth of structure, ultimately leading to a complete classification \cite{classification} and reconstruction \cite{reconstruction} of essentially generic solutions.

Somewhat surprisingly however, a complete proof of classical integrability of the superstring on $AdS_5 \times S^5$ was only achieved relatively recently, first in a `weak' sense in the pure spinor formalism by Sch\"afer-Nameki and Mikhailov \cite{SNM} and a year later in a `strong' sense in both the Green-Schwarz and pure spinor formalisms by Magro \cite{Magro}; the precise sense in which these respective results are `weak' and `strong' will be clarified. The reason it took so long to establish a proof is that the standard definition of classical integrability, in its modern incarnation \cite{BBT}, requires not only the existence of a Lax connection but also that the Poisson bracket of its spatial component takes on a very special form \cite{Maillet}. And yet all earlier attempts \cite{attempts} at proving this second point for the Bena-Polchinski-Roiban (BPR) connection were never entirely conclusive (it is worthwhile noting that the bosonic and principal chiral cases \cite{b_attempts} were not faced with the same difficulties). The reason for this difficulty was hinted at in \cite{Magro} where the calculation could be successfully brought to completion after judiciously amending the BPR connection with terms proportional to the Hamiltonian constraints.

The object of this paper is to justify the extension of the BPR connection used in \cite{Magro} as well as clarify its relation to the weaker result of \cite{SNM}. Now since the Hamiltonian formalism is better suited for the purpose of addressing the question of integrability, our approach will be to (re)build the spatial component of the Lax connection from scratch with a purely Hamiltonian mindset. In particular, since the superstring on $AdS_5 \times S^5$ has many gauge symmetries \---\ from world-sheet diffeomorphisms, $\kappa$-symmetry and the coset nature of the target space \---\ it will be important to correctly treat the corresponding constraints in the Hamiltonian setting \cite{Henneaux}.
Our definition of integrability in the presence of constraints will be the same as usual but with the additional requirement that it `sees' the constraints, in the sense that
\begin{itemize}
   \item[$(\text{I})$] The integrals of motion should all be first class.
   \item[$(\text{II})$] The zero curvature equation should hold strongly.
\end{itemize}
In other words, condition $(\text{II})$ requires that the Lax connection be flat in the whole of phase-space and not just on the constraint surface. This is the normal requirement for integrability in the absence of constaints. It is also very natural in the constrained case because prior to fixing any gauge it is desirable to make statements which hold in the entire phase-space and are not restricted to the constraint surface. Indeed, one might later be interested in treating the gauge invariance using BRST symmetry which requires extending the phase-space even further. The fact that we are dealing with a constrained system is handled by condition $(\text{I})$ which requires that all the `hidden' symmetries generated by the Lax connection leave the constraint surface invariant. This requirement is very natural since it effectively guarantees that we are indeed discussing the integrability of the constrained system.

Mimicking the strategy used in \cite{BPR} for constructing a flat connection, we will start from a general linear combination of the phase-space variables and require it to satisfy conditions $(\text{I})$ and $(\text{II})$ in turn. Condition $(\text{I})$ by itself will be enough to produce a weakly flat connection, namely the BPR connection with fermionic constraints added, which modulo ghost currents is equivalent to the pure spinor connection used in \cite{SNM}. Imposing condition $(\text{II})$ will add to this a further bosonic constraint to yield exactly the connection proposed in \cite{Magro}.

The paper is organised as follows. In section 1 we outline the general properties of integrable systems when constraints are present, motivating the conditions $(\text{I})$ and $(\text{II})$. We also illustrate our construction of a Lax matrix based on these conditions in the finite dimensional case. In section 2 we set up the Hamiltonian formalism for the Green-Schwarz superstring on $AdS_5 \times S^5$, using Dirac's consistency algorithm to identify all the relevant constraints. We then carry out the construction of the Lax connection by enforcing the conditions $(\text{I})$ and $(\text{II})$ in sections 3.1 and 3.2 respectively.

\begin{figure}[h]
\centering
\includegraphics[width=0.33\textwidth]{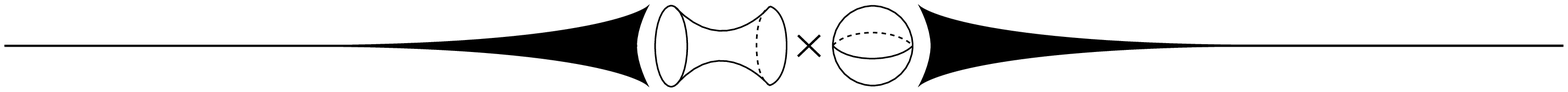}
\end{figure}

\vspace{-10mm}

\tableofcontents
\newpage

\section{Constrained integrable systems}

In the Hamiltonian scheme \cite{Henneaux}, a gauge system is restricted by a set of constraints $\{ \phi^A \approx 0 \}$ to live in a submanifold $\Sigma$ of the full phase-space $(\P, \{ \cdot, \cdot \})$. As usual `$\approx$' denotes equality on $\Sigma$. Furthermore, it is the first class property $\{ H, \phi^A \} \approx 0$ of the Hamiltonian $H$ which ensures that the system stays bound to the constraint surface $\Sigma$ at all times. To simplify the discussion, in this section we will treat the case of a finite-dimensional system, $\dim \P = 2 n$. 

\paragraph{Extensions.} Given two functions $F,G \in C^{\infty}(\P)$ whose restriction to the constraint surface $\Sigma$ are equal, \textit{i.e.} $F \approx G$, their difference can be written as a sum over all constraints $\phi^A$. Thus any $F \in C^{\infty}(\P)$ can be freely `extended' as $F \rightsquigarrow F + \sum_A f_A \phi^A$. When $F$ is first class it should only be extended using first class constraints $\{ \gamma^a \approx 0 \}$, with the property that $\{ \gamma^a, \phi^A \} \approx 0$. In the case of the Hamiltonian this leads to the introduction of the extended Hamiltonian $H_E = H + \sum_a u_a \gamma^a$ which combines the dynamics with arbitrary `gauge' transformations.

One can also extend any function, including first class ones, by terms not less than quadratic in the constraints since these have no effect at all on $\Sigma$. For instance we can modify the extended Hamiltonian as $H_E \rightsquigarrow H_E + \sum_{\alpha, \beta} u_{\alpha\beta} \chi^{\alpha} \chi^{\beta}$, where $\{ \chi^{\alpha} \approx 0 \}$ is the set of second class constraints, without affecting the dynamics on the constraint surface $\Sigma$.

\paragraph{Symmetry.} Consider a Hamiltonian action \cite{DaSilva} of $\mathbb{R}$ on $\P$ with moment map $\mu : \P \rightarrow \mathbb{R}$. For this action to be a symmetry of the constrained system it clearly ought to preserve the
\begin{wrapfigure}{r}{0.5\textwidth}
\vspace{-7mm}
\begin{center}
\psfrag{g}{\color{red} $\{ \cdot, \gamma^a \}$} \psfrag{p}{$\phi^A \approx 0$} \psfrag{f}{\color{blue} $\{ \cdot, \mu \}$}
\includegraphics[width=0.45\textwidth]{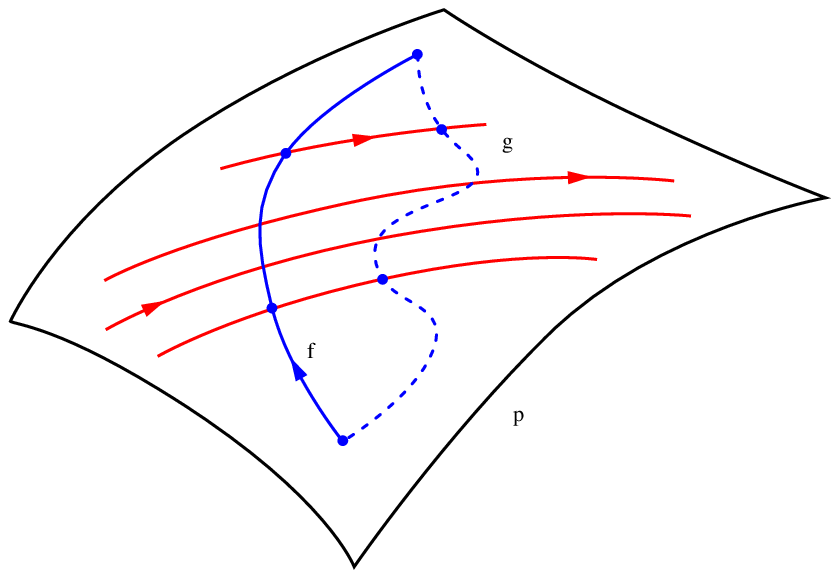}
\end{center}
\vspace{-7mm}
\end{wrapfigure}
constraint surface $\{ \phi^A \approx 0 \}$. Therefore the moment map should be first class
\begin{subequations}
\begin{equation} \label{sym first class}
\{ \mu, \phi^A \} \approx 0,
\end{equation}
after a possible extension $\mu \rightsquigarrow \mu + \sum_{\alpha} m_{\alpha} \chi^{\alpha}$. Furthermore, any extension of the type
\begin{equation} \label{sym ambiguity}
\mu \rightsquigarrow \mu + \sum_a m_a \gamma^a + \sum_{A,B} m_{AB} \phi^A \phi^B
\end{equation}
\end{subequations}
won't affect the first class property \eqref{sym first class} and has the effect of combining arbitrary gauge transformations with the actual symmetry, the quadratic terms having no effect at all on $\Sigma$. Such considerations can easily be generalised to Hamiltonian actions of a group $G$ on $\P$ with moment map $\mu : \P \rightarrow \gd$.

\paragraph{Integrability.} Loosely speaking an integrable system is one which admits as many symmetries as degrees of freedom. Let $p$ denote the number of first class constraints $\{ \gamma^a \}$ and $2m$ the number of second class constraints $\{ \chi^{\alpha} \}$. Since each first class constraint $\gamma^a$ generates a gauge symmetry on the constraint surface $\Sigma$, we are left with a total of $2(n-m-p) \equiv 2k$ independent phase-space variables. We will therefore say that the system is integrable if it admits a faithful Hamiltonian action of the $k$-torus $\mathbb{T}^k$ with moment map $I = (I_1, \ldots, I_k) : \P \rightarrow \mathbb{R}^k$ such that the Hamiltonian $H \in C^{\infty}(\P)$, which is unambiguously defined only on the constraint surface $\Sigma$, can be written weakly in terms of the moment map $H \approx H(I)$. The requirement that $I$ is a moment map in particular means that $\{ I_i \}$ should be independent almost everywhere and since the torus is abelian they must also be in involution,
\begin{equation} \label{involution}
\{ I_i, I_j \} = 0.
\end{equation}
As before the moment map should also be first class, which means that the integrals $\{ I_i \}$ must all be first class functions, \textit{i.e.} $\{ I_i, \phi^A \} \approx 0$, as in \eqref{sym first class}. The freedom \eqref{sym ambiguity} however is not available if we want to preserve \eqref{involution} although it might be needed in order to establish \eqref{involution}.

\paragraph{Lax matrix.} The modern formulation of integrable systems \cite{BBT} is based on the existence of a Lax pair $L, M$ of matrix valued functions on $\P$. In the constrained case these should \textit{a priori} have the property that the Hamiltonian equations on the constraint surface $\Sigma$ can be written in the weak Lax form
\begin{equation} \label{weak lax}
\{ L, H \} \approx [L, M],
\end{equation}
where $H$ is the Hamiltonian. However, we will demand that the integrals $I_i$ be obtainable from the spectral invariants $\tr L^j$ of the Lax matrix $L$, which could require a possible extension
\begin{equation} \label{lax extension}
L \rightsquigarrow L + \sum_A \lambda_A \phi^A.
\end{equation}
By a well known theorem of Babelon and Viallet \cite{Babelon-Viallet} the strong evolution property \eqref{involution} is then equivalent to the existence of an $r$-matrix such that the following strong equality holds,
\begin{equation} \label{r-matrix}
\{ L_{\tensor{1}}, L_{\tensor{2}} \} = [r_{\tensor{12}}, L_{\tensor{1}}] - [r_{\tensor{21}}, L_{\tensor{2}}].
\end{equation}
Taking the trace over the second tensor space yields $\{ L_{\tensor{1}}, \tr L^j \} = [ L_{\tensor{1}}, j \tr_{\tensor{2}} (r_{\tensor{12}} L^{j-1}_{\tensor{2}}) ]$, which immediately implies that by extending the Hamiltonian $H$ (and $M$) appropriately we can ensure that \eqref{weak lax} holds strongly
\begin{equation} \label{strong lax}
\{ L, H \} = [L, M].
\end{equation}
And more generally, every integral $I_i$ will have associated with it a Lax matrix $M_i$ such that $\{ L, I_i \} = [L, M_i]$. This reasoning also tells us that the extension of $H$ will simply be a spectral invariant of $L$ after the extension \eqref{lax extension}, typically a residue of $\tr L^2$.

\section{Green-Schwarz superstring on $AdS_5 \times S^5$}

Our starting point is the Lagrangian of the Green-Schwarz superstring on $AdS_5 \times S^5$ described as a supercoset model on $PSU(2,2|4)/SO(4,1) \times SO(5)$ \cite{Metsaev}
\begin{equation} \label{Lagrangian}
\mathcal{L} = -\frac{\sqrt{\lambda}}{2} \left[ \gamma^{\alpha \beta} \str \big(A_{\alpha}^{(2)} A_{\beta}^{(2)}\big) + \kappa \epsilon^{\alpha \beta} \str \big(A_{\alpha}^{(1)} A_{\beta}^{(3)}\big) \right] + \str \big( \Lambda (\partial_0 A_1 - \nabla_1 A_0) \big).
\end{equation}
where as usual $\nabla_1 \equiv \partial_1 - [A_1, \cdot]$ and the current $A = - g^{-1} dg = A^{(0)} + A^{(1)} + A^{(2)} + A^{(3)}$ takes values in the Grassmann envelope $\mathfrak{psu}(2,2|4; \Gamma)$ of the $\mathbb{Z}_4$-graded Lie superalgebra $\mathfrak{psu}(2,2|4)$. Here $\gamma^{\alpha \beta} = \sqrt{-h} h^{\alpha \beta}$ is the Weyl invariant combination of the worldsheet metric $h_{\alpha \beta}$ while $\epsilon^{\alpha \beta}$ is the two-dimensional anti-symmetric symbol and $\Lambda$ is a Lagrange multiplier imposing the Maurer-Cartan equation $dA = A^2$. In the following section we give a detailed account of the Hamiltonian analysis of \eqref{Lagrangian} using Dirac's procedure \cite{Henneaux} to determine all the constraints.

\subsection{Dirac's consistency algorithm}

\paragraph{Primary constraints.} The independent dynamical variables are $(A_0, A_1, \Lambda, h_{\alpha \beta})$ and their respective conjugate momenta are
\begin{equation*}
\Pi_0 \equiv \frac{\partial \L}{\partial \dot{A}_0}
\approx 0, \qquad \Pi_1 \equiv \frac{\partial
\L}{\partial \dot{A}_1} \approx \Lambda, \qquad
\Pi_{\Lambda} \equiv \frac{\partial \L}{\partial
\dot{\Lambda}} \approx 0, \qquad p^{\alpha \beta} \equiv
\frac{\partial \L}{\partial \dot{h}_{\alpha \beta}}
\approx 0.
\end{equation*}
Since none of these definitions depend on time derivatives each of them corresponds to primary constraints. Noting that $\Pi_{\Lambda}$ and $\Pi_1 - \Lambda$ form a set of second class constraints we eliminate them by introducing the corresponding Dirac bracket $\{,\}^{\ast}$. But for any pair of functions $F,G$ which do not depend on $\Pi_{\Lambda}$ one has $\{ F,G \}^{\ast} = \{ F,G \}$. We can therefore set $\Pi_1 - \Lambda = 0$ strongly to zero and keep the Poisson bracket. After doing this the Hamiltonian density reads
\begin{equation*}
\H = \frac{\sqrt{\lambda}}{2} \left[ \gamma^{\alpha \beta} \str
\big(A_{\alpha}^{(2)} A_{\beta}^{(2)}\big) + \kappa
\epsilon^{\alpha \beta} \str \big(A_{\alpha}^{(1)}
A_{\beta}^{(3)}\big) \right] + \str (\Pi_1 \nabla_1 A_0).
\end{equation*}

\paragraph{Secondary constraints.} The requirement that each remaining primary constraint be preserved under the Hamiltonian $H = \int d\sigma \H$ leads to a further set of secondary constraints. Thus $\dot{\Pi}_0 = \{ \Pi_0, H \} \approx 0$ yields for each grading the following constraints,
\begin{gather*}
\C^{(0)} = (\nabla_1 \Pi_1)^{(0)} \approx 0, \qquad
\C^{(2)} = - \gamma^{0 \alpha} A^{(2)}_{\alpha} + \frac{1}{\sqrt{\lambda}} (\nabla_1 \Pi_1)^{(2)} \approx 0,\\
\C^{(1)} = \frac{\sqrt{\lambda}}{2} \kappa A^{(1)}_1 +
(\nabla_1 \Pi_1)^{(1)} \approx 0, \qquad \C^{(3)} = -
\frac{\sqrt{\lambda}}{2} \kappa A^{(3)}_1 + (\nabla_1 \Pi_1)^{(3)}
\approx 0.
\end{gather*}
Since $\C^{(2)}$ is second class with the grade-$2$ part $\Pi^{(2)}_0$ of the primary constraint $\Pi_0$ we introduce a Dirac bracket, which on functions independent of $\Pi_0$ is equivalent to the Poisson bracket. We can therefore set $\C^{(2)} = 0$ strongly to zero and also eliminate any dependence on $A^{(2)}_0$. For such functions the Hamiltonian density can be rewritten up to boundary terms as
\begin{equation} \label{Hamiltonian}
\H = \lambda^+ T_+ + \lambda^- T_- - \str \Big( A^{(3)}_0 \C^{(1)} \Big) - \str \Big( A^{(1)}_0 \C^{(3)}
\Big) - \str \Big( A^{(0)}_0 \C^{(0)} \Big),
\end{equation}
where $T_{\pm} \equiv \sqrt{\lambda} \str (\A^{(2)}_{\pm} \A^{(2)}_{\pm})$ with $\A^{(2)}_{\pm} \equiv \frac{1}{2 \sqrt{\lambda}} \big( (\nabla_1 \Pi_1)^{(2)} \mp \sqrt{\lambda} A^{(2)}_1\big)$,
and we have introduced a new set of variables for the metric $h_{\alpha \beta}$,
\begin{equation} \label{metric var}
\lambda^{\pm} \equiv \frac{1 \pm \gamma_{01}}{\gamma_{11}} = \frac{\sqrt{-h} \pm h_{01}}{h_{11}}, \quad \xi = \ln h_{11}.
\end{equation}
The advantage of this parameterisation is that $\lambda^{\pm}$ are invariant under Weyl transformations $h_{\alpha \beta} \mapsto e^{\phi} h_{\alpha \beta}$ of the metric, whereas $\xi$ transforms as $\xi \mapsto \xi + \phi$. Thence the Hamiltonian \eqref{Hamiltonian} is explicitly Weyl invariant because it depends solely on the Weyl invariant functions $\lambda^{\pm}$.

Let us return to the last of the primary constraints $p^{\alpha \beta} \approx 0$ which now reads
\begin{equation*}
\pi^{\lambda}_{\pm} \equiv \frac{\partial \L}{\partial
\dot{\lambda}_{\pm}} \approx 0, \qquad \pi_{\xi} \equiv
\frac{\partial \L}{\partial \dot{\xi}} \approx 0.
\end{equation*}
The constraint $\pi_{\xi}$ generates Weyl transformations which can be fixed by choosing the gauge fixing condition $\xi \approx 0$. But since the variables $\xi$ and $\pi_{\xi}$ never appear in any physical variables we can simply discard them.
Requiring that the constraints $\pi^{\lambda}_{\pm}$ be preserved under the Hamiltonian, namely $\dot{\pi}^{\lambda}_{\pm} = \{ \pi^{\lambda}_{\pm}, H \} \approx 0$, gives rise to the Virasoro constraints
\begin{equation} \label{Vir}
T_{\pm} \approx 0.
\end{equation}
Adding to \eqref{Hamiltonian} a linear combination of all possible constraints $T_{\pm}$, $\pi_{\pm}^{\lambda}$, $\C^{(0,1,3)}$ and $\Pi_0^{(0,1,3)}$, with arbitrary coefficients $\rho^{\pm}$, $\nu_{\pm}$, $\mu^{(0,1,3)}$ and $\alpha^{(0,1,3)}$, gives the extended Hamiltonian density
\begin{equation} \label{Extended Hamiltonian}
\begin{split}
\H_E &= (\lambda^+ + \rho^+) T_+ + (\lambda^- + \rho^-)
T_- + \nu_+ \pi^{\lambda}_+ + \nu_- \pi^{\lambda}_-\\
&- \str \Big( \big(A^{(3)}_0 + \mu^{(3)}\big) \C^{(1)}
\Big) - \str \Big( \big(A^{(1)}_0 + \mu^{(1)}\big)
\C^{(3)} \Big) - \str \Big( \big(A^{(0)}_0 +
\mu^{(0)}\big) \C^{(0)} \Big)\\ &+ \str \big(\alpha^{(3)}
\Pi_0^{(1)}\big) + \str \big(\alpha^{(1)} \Pi_0^{(3)}\big) + \str
\big(\alpha^{(0)} \Pi_0^{(0)}\big).
\end{split}
\end{equation}

\paragraph{Tertiary constraints.} Next we must ensure that the new secondary constraints are preserved under time evolution. The simplest of these to consider is the constraint $\C^{(0)}$ for which one can straightforwardly show that $\dot{\C}^{(0)} = \{ \C^{(0)}, H_E \} \approx 0$.

The requirement that the constraints $\C^{(1,3)}$ be preserved is more involved. Recalling first that the Lagrangian \eqref{Lagrangian} is invariant under $\kappa$-symmetry only when $\kappa = \pm 1$ \cite{Arutyunov}, from now on we choose $\kappa = 1$. After some algebra, the conditions $\dot{\C}^{(1)} \approx \dot{\C}^{(3)} \approx 0$ boil down to
\begin{equation} \label{multipliers}
\left[ \A^{(2)}_+, \tilde{\mu}^{(1)} \right] \approx \left[ \A^{(2)}_-, \tilde{\mu}^{(3)} \right] \approx 0,
\end{equation}
where $\tilde{\mu}^{(1,3)}$ are related to $\mu^{(1,3)}$ through $A^{(1,3)}_0 + \mu^{(1,3)} = \pm (\lambda^{\pm} + \rho^{\pm}) A^{(1,3)}_1 + \tilde{\mu}^{(1,3)}$.
Thus the preservation of the secondary constraints $\C^{(1,3)}$ doesn't give rise to any tertiary constraints but rather to a restriction of their respective Lagrange multipliers $A^{(1,3)}_0 + \mu^{(1,3)}$.

Finally we turn to the preservation of the constraints $T_{\pm}$. One can show that,
\begin{equation*}
\dot{T}_+ \approx - 2 \sqrt{\lambda} \str \left( A^{(3)}_1 \big[ \A^{(2)}_-, A^{(3)}_0 + \mu^{(3)} \big] \right), \qquad
\dot{T}_- \approx 2 \sqrt{\lambda} \str \left( A^{(1)}_1 \big[ \A^{(2)}_+, A^{(1)}_0 + \mu^{(1)} \big] \right).
\end{equation*}
At first sight the conditions $\dot{T}_{\pm} \approx 0$ could impose further restrictions on the multipliers $\mu^{(1,3)}$. However, \eqref{multipliers} already yields $\dot{T}_{\pm} \approx 0$ so no further restriction on the Lagrange multipliers is required to preserve the secondary constraints $T_{\pm}$ and the consistency algorithm terminates.

Using the expression for $\mu^{(1,3)}$ in terms of $\tilde{\mu}^{(1,3)}$ one can reorganise the extended Hamiltonian density for the variables $\big( \lambda^{\pm}, \pi_{\pm}^{\lambda}, A_0^{(0,1,3)}, \Pi_0^{(0,1,3)}, A_1^{(0,1,2,3)}, \Pi_1^{(0,1,2,3)} \big)$ in the following way
\begin{multline} \label{Hamiltonian 2}
\H_E = (\lambda^+ + \rho^+) \T_+ + (\lambda^- + \rho^-) \T_- - \str \Big( \tilde{\mu}^{(3)}
\C^{(1)} \Big) - \str \Big( \tilde{\mu}^{(1)} \C^{(3)}
\Big) - \str \Big( \big( A^{(0)}_0 + \mu^{(0)} \big) \C^{(0)} \Big)\\
+ \str \big(\alpha^{(3)} \Pi_0^{(1)}\big) + \str \big(\alpha^{(1)}
\Pi_0^{(3)}\big) + \str \big(\alpha^{(0)} \Pi_0^{(0)}\big) + \nu_+
\pi^{\lambda}_+ + \nu_- \pi^{\lambda}_-,
\end{multline}
where we have added to the Virasoro constraints $T_{\pm}$ extra terms proportional to the fermionic constraints by defining the following `shifted' Virasoro constraints (\textit{c.f.} \cite{Hori}),
\begin{equation} \label{Vir 2}
\T_+ = T_+ - \str \left( A^{(1)}_1 \C^{(3)} \right), \qquad \T_- = T_- + \str \left( A^{(3)}_1 \C^{(1)} \right).
\end{equation}

\subsection{First and second class constraints}

To summarise the situation so far, we have a total of five primary constraints
\begin{equation*}
\varphi^1 \equiv \Pi^{(0)}_0 \approx 0, \quad \varphi^2 \equiv
\Pi^{(1)}_0 \approx 0, \quad \varphi^3 \equiv \Pi^{(3)}_0 \approx 0,
\qquad\qquad \varphi^{4,5} \equiv \pi^{\lambda}_{\pm} \approx 0,
\end{equation*}
whose respective stability gives rise to five more secondary constraints
\begin{gather*}
\varphi^6 \equiv \C^{(0)} \approx 0, \quad \varphi^7 \equiv
\C^{(1)} \approx 0, \quad \varphi^8 \equiv \C^{(3)}
\approx 0,\qquad\qquad \varphi^{9,10} \equiv \T_{\pm} \approx 0.
\end{gather*}
These in turn do not give rise to any tertiary constraints. The distinction between primary and secondary constraints however is of no real interest. A more relevant separation is that between first and second class constraints, which requires computing the various Poisson brackets between all the constraints $\{ \varphi^I \}_{I = 1}^{10}$.

It is obvious that all the primary constraints are first class since
\begin{equation*}
\{ \Pi^{(0)}_0, \varphi^I \} = \{ \Pi^{(1)}_0, \varphi^I \} = \{
\Pi^{(3)}_0, \varphi^I \} = \{ \pi^{\lambda}_{\pm}, \varphi^I \} = 0,
\qquad I = 1, \ldots, 10.
\end{equation*}
As for the second class constraints, one can show for instance that $\mathcal{C}^{(0)}$ is also first class since
\begin{equation*}
\{ \C^{(0)}_\1 (\sigma), \C^{(j)}_\2 (\sigma') \} = [C^{(00)}_{\1\2}, \C^{(j)}_\2 (\sigma)] \delta_{\sigma \sigma'} \approx 0, \quad j = 0,1,3,
\qquad \{ \C^{(0)} (\sigma), T_{\pm} (\sigma') \} = 0.
\end{equation*}
We refer to \cite{Magro} for a definition of the quadratic Casimir $C_{\1\2}$ and its different gradings $C^{(i \, 4-i)}_{\1\2}$. As usual the Virasoro constraints form a closed algebra \cite{Brink-Henneaux},
\begin{equation*}
\{ T_{\pm}(\sigma) , T_{\pm}(\sigma') \} = \mp \left[ T_{\pm}(\sigma) + T_{\pm}(\sigma') \right] \partial_{\sigma} \delta_{\sigma \sigma'} \approx 0, \qquad \{ T_+(\sigma) , T_-(\sigma') \} = 0.
\end{equation*}
On the other hand the algebra of the fermionic constraints $\C^{(1,3)}$ does not close, indicating the fact that they are partly second class. Next, computing the bracket between the Virasoro constraints $T_{\pm}$ and the $\C^{(1,3)}$ reveals that they do not commute. This is slightly puzzling at first since one would expect the generators of conformal symmetry to be first class \cite{Brink-Henneaux}. However one recalls that the Virasoro constraints ought to be shifted as in \eqref{Vir 2}. And indeed one can show that the brackets between the constraints $\T_{\pm}$ and $\C^{(1,3)}$ all weakly vanish,
\begin{equation*}
\{ \T_{\pm}(\sigma), \C^{(1)}(\sigma') \} \approx 0, \qquad \{ \T_{\pm}(\sigma), \C^{(3)}(\sigma') \} \approx 0.
\end{equation*}
The upshot is that the true generators of conformal symmetry are not the Virasoro constraints $T_{\pm}$ themselves but rather their shifts $\T_{\pm}$ defined in \eqref{Vir 2}. One can check that they too satisfy the Virasoro algebra, namely
\begin{equation*}
\{ \T_{\pm}(\sigma) , \T_{\pm}(\sigma') \} = \mp \left[ \T_{\pm}(\sigma) + \T_{\pm}(\sigma') \right]
\partial_{\sigma} \delta_{\sigma \sigma'}, \qquad \{ \T_+(\sigma) , \T_-(\sigma') \} = 0.
\end{equation*}
In conclusion all the primary and secondary constraints $\{ \varphi^I \}_{I=1}^{10}$ are first class \textit{except} for the fermionic constraints $\varphi^7 = \C^{(1)}$ and $\varphi^8 = \C^{(3)}$ which are partly second class since they fail to commute amongst themselves, \textit{i.e.} $\{ \C^{(1)}, \C^{(3)} \} \not \approx 0$.

\subsection{Partial gauge fixing}

At this stage it is possible to completely fix the gauge freedom generated by each one of the primary constraints $\pi_{\pm}^{\lambda}$ and $\Pi_0^{(0,1,3)}$. For this we choose the following set of contraints,
\begin{equation} \label{partial gauge}
\begin{split}
c_{\pm} &\equiv \lambda^{\pm} - 1 \approx 0, \qquad \quad \D^{(1)} \equiv A_0^{(1)} - \lambda^- A_1^{(1)} \approx 0,\\
\D^{(0)} &\equiv A_0^{(0)} - A^{(0)}_1 \approx 0, \qquad \D^{(3)} \equiv A_0^{(3)} + \lambda^+ A_1^{(3)} \approx 0.
\end{split}
\end{equation}
That these are a good set of gauge fixing conditions follows from the fact that $c_{\pm} \approx \D^{(0,3,1)} \approx 0$ are second class with $\pi_{\pm}^{\lambda} \approx \Pi_0^{(0,1,3)} \approx 0$ respectively because
\begin{gather*}
\{ c_{\pm}, \pi_{\pm}^{\lambda} \} = 1, \quad \{ \D^{(3)}, \pi_+^{\lambda} \} = A_1^{(3)} \delta_{\sigma \sigma'}, \quad \{ \D^{(1)}, \pi_-^{\lambda} \} = -A_1^{(1)} \delta_{\sigma \sigma'},\\ \{ \D_\2^{(0)}, \Pi_{0\1}^{(0)} \} = C_{\1\2}^{(00)} \delta_{\sigma \sigma'}, \quad \{ \D_\2^{(3)}, \Pi_{0\1}^{(1)} \} = C_{\1\2}^{(13)} \delta_{\sigma \sigma'}, \quad \{ \D_\2^{(1)}, \Pi_{0\1}^{(3)} \} = C_{\tensor{12}}^{(31)} \delta_{\sigma \sigma'}.
\end{gather*}
Restricting attention to functions of the variables $(A_1^{(0,1,2,3)}, \Pi_1^{(0,1,2,3)})$ their Dirac and Poisson brackets coincide. We can therefore set all the constraints $c_{\pm} = \D^{(0,1,3)} = \pi_{\pm}^{\lambda} = \Pi_0^{(0,1,3)} = 0$ strongly to zero and the extended Hamiltonian becomes
\begin{equation} \label{Hamiltonian 3}
\H_E = \rho^+ \T_+ + \rho^- \T_- - \str \Big( \tilde{\mu}^{(3)} \C^{(1)} \Big) - \str \Big( \tilde{\mu}^{(1)} \C^{(3)} \Big) - \str \Big( \mu^{(0)} \C^{(0)} \Big),
\end{equation}
where the definitions of the variables $\rho^{\pm}$ have both been shifted by $-1$ and $\mu^{(0)}$ by $-A^{(0)}_1$.

By working in a particular representation of $\mathfrak{su}(2,2|4)$ where $\A^{(2)}_{\pm} \A^{(2)}_{\pm} \approx c \cdot {\bf 1}$ for some $c$ (see for instance \cite{Arutyunov}), we can solve the constraints \eqref{multipliers} on the multipliers $\tilde{\mu}^{(1,3)}$ as
\begin{equation*}
\tilde{\mu}^{(1)} \approx 2 \sqrt{\lambda} [\A^{(2)}_+, i k^{(1)}]_+, \qquad \tilde{\mu}^{(3)} \approx 2 \sqrt{\lambda} [\A^{(2)}_-, i k^{(3)}]_+,
\end{equation*}
where $k^{(1,3)}$ are completely arbitrary. This then allows us to write the extended Hamiltonian \eqref{Hamiltonian 3} solely in terms of first class parts of the fermionic constraints $\C^{(1,3)}$, namely
\begin{equation} \label{K13}
\K^{(1)} \equiv 2 \sqrt{\lambda} [\A^{(2)}_-, i\C^{(1)}]_+, \qquad \K^{(3)} \equiv 2 \sqrt{\lambda} [\A^{(2)}_+, i\C^{(3)}]_+.
\end{equation}

In conclusion of the above analysis, the setup for the Hamiltonian dynamics of the Green-Schwarz superstring on $AdS_5 \times S^5$ comprises the following ingredients:
\begin{subequations} \label{setup}
\begin{itemize}
    \item  The total phase-space $\P$ is parametrised by the coordinates $(A_1^{(0,1,2,3)}, \Pi_1^{(0,1,2,3)})$ and equipped with the canonical Poisson structure
\begin{equation} \label{canon PB}
\{ A^{(i)}_{1\tensor{1}}(\sigma), \Pi^{(4-i)}_{1\tensor{2}}(\sigma') \} = C^{(i\, 4-i)}_{\tensor{12}} \delta(\sigma - \sigma').
\end{equation}
    \item  The system is subject to the bosonic constraints $\T_{\pm} \approx \C^{(0)} \approx 0$, all of which are first class. The fermionic constraints $\C^{(1,3)} \approx 0$ are partly first and second class, with their first class parts $\K^{(1,3)} \approx 0$ given in \eqref{K13}. We shall denote the set of all constraints as
\begin{equation}  \label{all constraints}
\{ \Phi^A \} \equiv \{ T_{\pm}, \C^{(0,1,3)} \}.
\end{equation}
Likewise the set of all fist class constraints will be denoted
\begin{equation} \label{FC constraints}
\{ \Gamma^a \} \equiv \{ \T_{\pm}, \C^{(0)}, \K^{(1,3)} \}.
\end{equation}
    \item  The extended Hamiltonian is the general linear combination of all the first class constraints,
\begin{equation} \label{Hamiltonian 4}
\H_E = \rho^+ \T_+ + \rho^- \T_- - \str \Big( k^{(3)} \K^{(1)} \Big) - \str \Big( k^{(1)} \K^{(3)} \Big) - \str \Big( \mu^{(0)} \C^{(0)} \Big).
\end{equation}
\end{itemize}
\end{subequations}

\section{Hamiltonian Lax connection}

The goal of this section is to construct the spatial component of the Lax connection within the Hamiltonian framework laid out in the previous section. In light of the strategy employed by Bena, Polchinski and Roiban in the Lagrangian framework \cite{BPR}, we consider a general linear combination of the phase-space variables
\begin{equation} \label{lax}
L = A_1^{(0)} + a A_1^{(1)} + b A_1^{(2)} + c A_1^{(3)} + \rho (\nabla_1 \Pi_1)^{(0)} + \gamma (\nabla_1 \Pi_1)^{(1)} + \beta (\nabla_1 \Pi_1)^{(2)} + \alpha (\nabla_1 \Pi_1)^{(3)},
\end{equation}
where the parameters $a, b, c, \alpha, \beta, \gamma, \rho$ are assumed constant for simplicity. To fix them we impose on $L$ the two fundamental requirements $(\text{I})$ and $(\text{II})$ stated in the introduction.

\subsection{A first class monodromy}

In a two-dimensional integrable field theory with Lax matrix $L$, the integrals of motion are given by the spectral invariants $\str \Omega(L)^n$ of the monodromy matrix $\Omega(L)$ \cite{BBT}. In the presence of constraints it is therefore necessary to ensure that the functions $\str \Omega(L)^n$ are first class, which indirectly imposes a restriction on the Lax matrix $L$.

The Poisson bracket of \eqref{lax} with a generic constraint $\Phi(\sigma)$ from \eqref{all constraints} takes the form
\begin{equation} \label{lax-phi}
\{ L_\1 (\sigma), \Phi_\2 (\sigma') \} = \eta X_{\1\2}(\sigma') \partial_{\sigma} \delta_{\sigma \sigma'} + [X_{\1\2}(\sigma), Y_\1(\sigma)] \delta_{\sigma \sigma'},
\end{equation}
where $X(\sigma), Y(\sigma)$ and $\eta$ are determined from $\Phi(\sigma)$. The resulting bracket for the transition matrix $T(\sigma_1, \sigma_2) = P \overleftarrow{\exp} \int_{\sigma_2}^{\sigma_1} d\sigma L(\sigma)$ reads
\begin{equation} \label{transition}
\begin{split}
\{ T_{\tensor{1}}(\sigma_1, \sigma_2), \Phi_{\tensor{2}}(\sigma')  \} &= \chi(\sigma'; \sigma_1, \sigma_2) T_{\tensor{1}}(\sigma_1, \sigma') \left( [ X_{\tensor{12}}(\sigma'), Y_{\tensor{1}}(\sigma') - \eta L_{\tensor{1}}(\sigma') ] \right) T_{\tensor{1}}(\sigma', \sigma_2)\\
&+ \eta X_{\tensor{12}}(\sigma_1) T_{\tensor{1}}(\sigma_1, \sigma_2) \delta'(\sigma' - \sigma_1) - \eta T_{\tensor{1}}(\sigma_1, \sigma_2) X_{\tensor{12}}(\sigma_2) \delta'(\sigma' - \sigma_2),
\end{split}
\end{equation}
where $\chi(\,\cdot\, ; \sigma_1, \sigma_2)$ is the indicator function for the interval $[\sigma_1,\sigma_2]$. Let $\Omega(L) = T(2 \pi, 0)$ denote the monodromy matrix associated with $L$. The key observation is that for the spectral invariants $\str \Omega(L)^n$ of the monodromy to weakly commute with the constraint $\Phi(\sigma)$ it is sufficient to require
\begin{equation} \label{pre zero curvature}
[X_{\1\2}, Y_\1 - \eta L_\1] \approx 0.
\end{equation}
Indeed, by killing the first term on the right hand side of \eqref{transition}  this condition leads to a simple bracket between $\Omega(L)$ and the modes $\Phi^{\varphi} = \int_0^{2 \pi} d\sigma' \str \varphi(\sigma') \Phi(\sigma')$ of $\Phi(\sigma')$, namely
\begin{equation} \label{FC monodromy}
\{ \Omega(L), \Phi^{\varphi}(\sigma')  \} \approx [ \Omega(L), X^{\varphi} ]
\end{equation}
where $X^{\varphi} \equiv \eta \str_{\tensor{2}} X_{\tensor{12}}(0) \varphi'_{\tensor{2}}(0)$. In turn, equation \eqref{FC monodromy} immediately implies that $\str \Omega(L)^n$ is first class for any $n \in \mathbb{N}$.

\paragraph{Fixing the parameters.} Choosing $\Phi$ among the constraints in \eqref{all constraints}, equation \eqref{pre zero curvature} yields a set of conditions to be imposed on the parameters of \eqref{lax}:
\begin{subequations} \label{fix coeffs}
\begin{itemize}
   \item With $\Phi = T_{\pm}$ we find $X_{\pm} = - \ha \A^{(2)}_{\pm}$, $\eta_{\pm} = b \mp \beta \sqrt{\lambda}$ and
\begin{equation*}
Y_{\pm} = \eta_{\pm} A^{(0)}_1 + \big(c - \alpha \sqrt{\lambda} (\ha \pm 1)\big) A^{(1)}_1 + A^{(2)}_1 + \big(a + \gamma \sqrt{\lambda} (\ha \mp 1)\big) A^{(3)}_1 + \ha \beta (\nabla_1 \Pi_1)^{(0)}.
\end{equation*}
Comparing this with the Lax matrix \eqref{lax} it is straightforward to show that the difference $Y_{\pm} - \eta_{\pm} L$ is weakly proportional to $\A^{(2)}_{\pm}$ provided
\begin{equation} \label{fix coeffs 1}
\eta^1 - 2 \alpha \sqrt{\lambda} = \eta_+ \eta^3, \quad \eta^1 = \eta_- \eta^3, \quad \eta_+ \eta_- = 1, \quad \eta^3 + 2 \gamma \sqrt{\lambda} = \eta_- \eta^1,
\end{equation}
where $\eta^1 = c + \alpha \ha \sqrt{\lambda}$ and $\eta^3 = a - \gamma \ha \sqrt{\lambda}$.
   \item With $\Phi = \C^{(1,3)}$ we find respectively $X^1_{\tensor{12}} = - C^{(13)}_{\tensor{12}}, X^3_{\tensor{12}} = - C^{(31)}_{\tensor{12}}$ and
\begin{align*}
Y^1 &= \eta^1 A^{(0)}_1 + A^{(1)}_1 + (\eta^3 + \gamma \sqrt{\lambda}) A^{(2)}_1 + \eta_- A^{(3)}_1 + \gamma (\nabla_1 \Pi_1)^{(2)},\\
Y^3 &= \eta^3 A^{(0)}_1 + \eta_+ A^{(1)}_1 + (\eta^1 - \alpha \sqrt{\lambda}) A^{(2)}_1 + A^{(3)}_1 + \alpha (\nabla_1 \Pi_1)^{(2)}.
\end{align*}
This time we need the differences $Y^{1,3} - \eta^{1,3} L$ to weakly vanish. This requirement only leads to one further condition on top of \eqref{fix coeffs 1}, namely
\begin{equation} \label{fix coeffs 2}
\eta^1 \eta^3 = 1.
\end{equation}
   \item With $\Phi = \C^{(0)}$ we find $X_{\tensor{12}} = - C^{(00)}_{\tensor{12}}$, $\eta = 1$ and $Y = L$ so that trivially $Y - \eta L = 0$.
\end{itemize}
\end{subequations}

Note that the conditions \eqref{fix coeffs} do not depend on $\rho$. They constitute a total of five independent constraints on the remaining six coefficients $\{ a, b, c, \alpha, \beta, \gamma \}$ which can therefore be parametrised by a single complex number $z$. One such solution exactly corresponds to the one put forward in \cite{Magro}, namely
\begin{subequations} \label{parameters}
\begin{alignat}{3}
a &= \frac{1}{4} (z^{-3} + 3 z),\qquad && b = \frac{1}{2} (z^{-2} + z^2),\qquad && c = \frac{1}{4} (3 z^{-1} + z^3),\\
\gamma &= \frac{1}{2 \sqrt{\lambda}} (z^{-3}-z),\qquad && \beta = \frac{1}{2 \sqrt{\lambda}} (z^{-2} - z^2),\qquad && \alpha = \frac{1}{2 \sqrt{\lambda}} (z^{-1} - z^3).
\end{alignat}
\end{subequations}
All other solutions are generated by applying $z \mapsto i z$ or $z \mapsto z^{-1}$. Plugging this parametrisation into the original Lax matrix \eqref{lax} and temporarily setting $\rho = 0$ it reads
\begin{equation} \label{lax H}
\begin{split}
L_{\text{H}} = L_{\text{BPR}} + \frac{1}{2 \sqrt{\lambda}}(z^{-3} - z) \C^{(1)} + \frac{1}{2 \sqrt{\lambda}}(z^{-1} - z^3) \C^{(3)}
\end{split}
\end{equation}
where $L_{\text{BPR}}$ is the (spatial component of the) standard Bena-Polchinski-Roiban (BPR) Lax connection \cite{BPR} expressed in terms of phase-space variables, which can be written as
\begin{equation} \label{lax BPR}
\begin{split}
L_{\text{BPR}} =  z^{-2} \A_-^{(2)} + z^{-1} A_1^{(3)} + A_1^{(0)} + z A_1^{(1)} - z^2 \A_+^{(2)}.
\end{split}
\end{equation}
An important remark is in order. Had we only insisted on imposing the first four conditions \eqref{fix coeffs 1} coming from the bosonic constraints $T_{\pm}$, which would have been justified if dealing with the bosonic string, the set of conditions on the parameters of the Lax matrix \eqref{lax} would have also admitted the BPR connection as a solution (namely $a, c, \alpha, \gamma = 0$ and $b, \beta$ as in \eqref{parameters}). In other words, to correctly take into account the effect of fermions and $\kappa$-symmetry we must also impose the extra condition \eqref{fix coeffs 2} coming from the fermionic constraints $\C^{(1,3)}$.

\paragraph{Admissible Lax matrices.} Although the two connections $L_{\text{H}}$ and $L_{\text{BPR}}$ agree on the constraint surface, by construction $L_H$ has the desirable property that all the spectral invariants of its monodromy matrix $\Omega(L_{\text{H}})$ are first class. Yet this statement fails to be true for $\Omega(L_{\text{BPR}})$ and this difference is due to the second class nature of the fermionic constraints $\C^{(1, 3)}$. On the other hand, adding an arbitrary linear combination of the first class constraints to $L_H$ doesn't destroy its special property, so that $\rho \, \C^{(0)}$ can be reinstated with $\rho$ arbitrary. In fact, at this stage nothing prevents us from adding an arbitrary linear combination of the first class constraints to \eqref{lax H}, leading to the most general admissible Lax matrix, with $\alpha_a \in C^{\infty}(\P)$,
\begin{equation} \label{lax general}
L = L_{\text{H}} + \sum_a \alpha_a \Gamma^a.
\end{equation}

\subsection{Strong zero curvature equation}

We will now fix the first class part of the Lax connection \eqref{lax general} such that it becomes the spatial component of a strongly flat connection. For simplicity we shall go back to \eqref{lax} and limit our search to the following restricted class of admissible Lax matrices,
\begin{equation} \label{L_H + rho}
L = L_{\text{H}} + \rho\, \C^{(0)}.
\end{equation}
But in order to write down a zero curvature equation we first need to identify momentum and energy since these generate translations in $\sigma$ and $\tau$ respectively.

\paragraph{Momentum.} The generator of $\sigma$-translations is given by the density
\begin{equation} \label{mom density def}
\P_1 \equiv \T_+ - \T_- - \str(A^{(0)}_1 \C^{(0)}),
\end{equation}
in the sense that $\{ F(\sigma), \P_1(\sigma') \} = F(\sigma') \partial_{\sigma} \delta_{\sigma \sigma'}$ or equivalently
\begin{equation} \label{mom density eq}
\{ F(\sigma), P_1 \} = \partial_{\sigma} F(\sigma)
\end{equation}
where $P_1 = \int d\sigma' \P_1(\sigma')$ is the total momentum. This is easy to show for the different gradings of the canonical variables $A_1$, $\nabla_1 \Pi_1$ and extends by Leibniz's rule to any other function $F$. Note that it is the shifted Virasoro generators $\T_{\pm}$ which appear in \eqref{mom density def} rather than $T_{\pm}$. Moreover, a further extension by the constraint $\C^{(0)}$ is required for \eqref{mom density eq} to hold. This corresponds to the choice of non-zero rigid parameters $\rho^+ = -\rho^- = 1$, $\mu^{(0)} = A^{(0)}_1$ in the extended Hamiltonian \eqref{Hamiltonian 4}. We thus have for the general admissible Lax connection
\begin{equation*}
\{ L(\sigma, z), \P_1(\sigma') \} = L(\sigma', z) \partial_{\sigma} \delta_{\sigma \sigma'}.
\end{equation*}

\paragraph{Energy.} The question of $\tau$-translations is more delicate since $\tau$ is not intrinsically defined as opposed to the $\sigma$ variable. Indeed, the $\tau$-flow should correspond to the Lagrangian equations of motion on the constraint surface, but this requirement does not uniquely fix the generator of $\tau$-translations strongly since one can always add a quadratic combination of the constraints.

Let us start by considering the equations of motion on the constraint surface. In the partial gauge \eqref{partial gauge} the different gradings of the Maurer-Cartan equation $\partial_0 A_1 \approx \nabla_1 A_0$ take the form
\begin{subequations} \label{eom}
\begin{align}
\partial_0 A^{(0)}_1 &\approx \partial_1 A^{(0)}_1 + 2 [A^{(1)}_1, A^{(3)}_1] + 2 [A^{(2)}_1, \A^{(2)}_{\pm}],\\
\partial_0 A^{(1)}_1 &\approx \partial_1 A^{(1)}_1 + 2 [A^{(3)}_1, \A^{(2)}_+],\\
\sqrt{\lambda} \, \partial_0 A^{(2)}_1 &\approx - \partial_1 (\nabla_1 \Pi_1)^{(2)} + 2 \sqrt{\lambda} \, [A^{(0)}_1, \A^{(2)}_-],\\
\partial_0 A^{(3)}_1 &\approx - \partial_1 A^{(3)}_1 + 2 [A^{(0)}_1, A^{(3)}_1] + 2 [A^{(1)}_1, \A^{(2)}_-],
\end{align}
As for the bosonic Lagrangian equation of motion they read
\begin{equation}
\partial_0 (\nabla_1 \Pi_1)^{(2)} \approx - \sqrt{\lambda} \, \partial_1 A^{(2)}_1 + 2 \sqrt{\lambda} \, [A^{(0)}_1, \A^{(2)}_-].
\end{equation}
\end{subequations}
The fermionic Lagrangian equations of motion on the other hand are automatic in the partial gauge $\D^{(1,3)} \approx 0$ of \eqref{partial gauge}. The equations \eqref{eom} can all be written in the Hamiltonian form
\begin{equation*}
\partial_0 A^{(i)}_1 \approx \{ A^{(i)}_1, P_0 \}, \qquad \partial_0 (\nabla_1 \Pi_1)^{(2)} \approx \{ (\nabla_1 \Pi_1)^{(2)}, P_0 \}
\end{equation*}
where $P_0 \equiv \int d\sigma' \P_0(\sigma')$ and the energy density $\P_0$ is chosen as
\begin{equation} \label{pre energy def}
\P_0 = \T_+ + \T_- - \str (A^{(0)}_1 \C^{(0)}).
\end{equation}
Any attempt at finding a Lax connection which is flat in the strong sense using \eqref{pre energy def} as our definition of the energy will ultimately fail. As alluded to above the reason for this is that the equations of motion on the constraint surface can only determine the energy up to terms not less than quadratic in the constraints. Exploiting this freedom we can therefore add quadratic terms to \eqref{pre energy def} without affecting the induced flow on the constraint surface. We claim that the relevant extension of the energy is given by
\begin{equation} \label{energy extension}
\P_0 \rightsquigarrow \P_0 + \frac{1}{\sqrt{\lambda}} \str (\C^{(1)} \C^{(3)}).
\end{equation}
At the end of this section we will explain how such an extension can be ascertained. For the moment let us take \eqref{energy extension} for granted and introduce the generators of $\sigma^{\pm}$-translations
\begin{subequations} \label{P+ and P-}
\begin{align}
\P_+ &\equiv \T_+ + \frac{1}{2 \sqrt{\lambda}} \str (\C^{(1)} \C^{(3)}) - \str (A^{(0)}_1 \C^{(0)}),\\
\P_- &\equiv \T_- + \frac{1}{2 \sqrt{\lambda}} \str (\C^{(1)} \C^{(3)}).
\end{align}
\end{subequations}
In terms of these, the energy density is given by $\P_0 = \P_+ + \P_-$, and likewise the momentum density \eqref{mom density def} reads $\P_1 = \P_+ - \P_-$.

\paragraph{Fixing $\rho$.} With these definitions in place, we claim that provided
\begin{equation} \label{rho}
\rho = \frac{1}{2 \sqrt{\lambda}} (1-z^4),
\end{equation}
the flows generated by $\P_{\pm}$ in \eqref{P+ and P-} are strongly flat, in the sense that there exists respective Lax matrices $L_{\pm}$ with the property that $\{ L, \P_{\pm} \} = L_{\pm} \partial_{\sigma} \delta_{\sigma \sigma'} + [L_{\pm}, L] \delta_{\sigma \sigma'}$, or in other words
\begin{equation} \label{flat P+-}
\{ L, P_{\pm} \} = \partial_{\sigma} L_{\pm} + [L_{\pm}, L]
\end{equation}
where $P_{\pm} \equiv \int d\sigma' \P_{\pm}(\sigma')$. To show this one simply evaluates the bracket $\{ L, \P_{\pm} \}$ which takes the general form $A \, \partial_{\sigma} \delta_{\sigma \sigma'} + B \, \delta_{\sigma \sigma'}$. Computing the difference $B - [A,L]$ we find it is proportional to $1 - z^4 - 2 \sqrt{\lambda} \rho$ so that equation \eqref{flat P+-} holds if and only if $\rho$ is given by \eqref{rho}.

\paragraph{Strong zero curvature.} To summarise the result, let us denote the action of $P_0$ on the entire phase-space as $\partial_{\tau} \equiv \{ \cdot , P_0 \}$ and introduce the Lax matrix $L_0 = L_+ + L_-$. It then follows from \eqref{flat P+-} that the Lax matrix of \cite{Magro}
\begin{equation} \label{lax Marc}
L_{\text{M}} = L_{\text{BPR}} + \frac{1}{2 \sqrt{\lambda}} (1 - z^4) \left( \C^{(0)} + z^{-3} \C^{(1)} + z^{-1} \C^{(3)} \right)
\end{equation}
satisfies the following strong zero curvature equation,
\begin{equation} \label{strong zero curv}
[\partial_{\tau} - L_0, \partial_{\sigma} - L_{\text{M}}] = 0.
\end{equation}

Alternatively, the Lax matrices of $\sigma^{\pm}$-translations are explicitly given by
\begin{subequations} \label{L+-}
\begin{align}
L_+ &= A^{(0)}_1 + \frac{3}{4} z A^{(1)}_1 + \frac{1}{2} z^2 A^{(2)}_1 + \frac{1}{4} z^3 A^{(3)}_1 \notag\\
&\qquad\qquad - \frac{1}{2 \sqrt{\lambda}} \left( z (\nabla_1 \Pi_1)^{(1)} + z^2 (\nabla_1 \Pi_1)^{(2)} + z^3 (\nabla_1 \Pi_1)^{(3)} + (z^4 - 1) (\nabla_1 \Pi_1)^{(0)} \right),\\
L_- &= - \frac{1}{4} z^{-3} A^{(1)}_1 - \frac{1}{2} z^{-2} A^{(2)}_1 - \frac{3}{4} z^{-1} A^{(3)}_1 \notag\\
&\qquad\qquad - \frac{1}{2 \sqrt{\lambda}}  \left( z^{-3} (\nabla_1 \Pi_1)^{(1)} + z^{-2} (\nabla_1 \Pi_1)^{(2)} + z^{-1} (\nabla_1 \Pi_1)^{(3)} \right).
\end{align}
\end{subequations}
In particular, $L_+ \in \mathfrak{psu}(2,2|4)[z]$ is localised around $z=0$ whereas  $L_- \in \mathfrak{psu}(2,2|4)[z^{-1}]$ is localised around $z = \infty$. If we introduce the notation $\partial_{\pm} \equiv \{ \cdot , P_{\pm} \}$ for the derivatives in $\sigma^{\pm}$ respectively then \eqref{L+-} satisfy the following strong zero curvature equation,
\begin{equation} \label{lax L+-}
[\partial_+ - L_+, \partial_- - L_-] = 0.
\end{equation}

\paragraph{Strong $r/s$-matrix algebra.} Let us motivate the extension \eqref{energy extension} of the energy density $\P_0$. The Poisson bracket of \eqref{lax Marc} with itself was computed in \cite{Magro} and reads
\begin{equation} \label{r-s alg}
\{ L_{\tensor{1}}, L_{\tensor{2}} \} = [r_{\tensor{12}} - s_{\tensor{12}}, L_{\tensor{1}}] \delta_{\sigma \sigma'} + [r_{\tensor{12}} + s_{\tensor{12}}, L_{\tensor{2}}] \delta_{\sigma \sigma'} - 2 s_{\tensor{12}} \delta'_{\sigma \sigma'},
\end{equation}
where the tensor indices $\tensor{1}$ and $\tensor{2}$ respectively imply a dependence on $(\sigma,z)$ and $(\sigma', z')$. See for instance \cite{Maillet} for a general discussion of algebras of the form \eqref{r-s alg}. After multiplying by $2 L_{\tensor{1}}$ and taking the supertrace $\str_{\tensor{1}}$, equation \eqref{r-s alg} starts to resemble a zero curvature equation. Now since the energy and momentum ought to be extractable from the Lax matrix at its singular points $z = 0, \infty$ \cite{classification, reconstruction}, let us introduce the following differential
\begin{equation*}
\omega \equiv \frac{\sqrt{\lambda}}{16} \str L^2 du,
\end{equation*}
where $u = 2 \frac{1 + z^4}{1 - z^4}$ is the Zhukovsky variable. Although this definition depends quite naturally on $u$, since the Lax matrix is expressed in the $z$-variable we should rewrite $du$ in terms of $dz$. We can then write the densities \eqref{P+ and P-} simply as
\begin{equation} \label{hatP+-}
\P_+ = - \text{res}_{z = \infty} \, \omega, \quad \P_- = \text{res}_{z = 0} \, \omega.
\end{equation}

\paragraph{Weak $r/s$-matrix algebra.} Let us also compute the Poisson bracket of $L_{\text{H}} = L - \rho\, \C^{(0)}$ with itself starting from \eqref{r-s alg}. For this we need the Poisson brackets
\begin{equation*}
\{ L_{\text{H}\tensor{1}}, \C^{(0)}_{\tensor{2}} \} = - C^{(00)}_{\tensor{12}} \partial_{\sigma} \delta_{\sigma \sigma'} - [C^{(00)}_{\tensor{12}}, L_{\text{H}\tensor{1}}] \delta_{\sigma \sigma'}, \qquad
\{ \C^{(0)}_{\tensor{1}}, L_{\text{H}\tensor{2}} \} = - C^{(00)}_{\tensor{12}} \partial_{\sigma} \delta_{\sigma \sigma'} + [C^{(00)}_{\tensor{12}}, L_{\text{H}\tensor{2}}] \delta_{\sigma \sigma'}
\end{equation*}
as well as $\{ \C^{(0)}_{\tensor{1}}, \C^{(0)}_{\tensor{2}} \} = - [C^{(00)}_{\tensor{12}}, \C^{(0)}_{\tensor{1}}] \delta_{\sigma \sigma'}$. A short calculation shows that
\begin{equation*}
\begin{split}
\{ L_{\text{H}\tensor{1}}, L_{\text{H}\tensor{2}} \} = \; &[r_{\tensor{12}} - s_{\tensor{12}} + \rho_2 C^{(00)}_{\tensor{12}}, L_{\text{H}\tensor{1}}] \delta_{\sigma \sigma'} + [r_{\tensor{12}} + s_{\tensor{12}} - \rho_1 C^{(00)}_{\tensor{12}}, L_{\text{H}\tensor{2}}] \delta_{\sigma \sigma'}\\
&- (2 s_{\tensor{12}} - (\rho_1 + \rho_2) C^{(00)}_{\tensor{12}}) \partial_{\sigma} \delta_{\sigma \sigma'} + [\rho_1 \rho_2 C^{(00)}_{\tensor{12}}, \C^{(0)}_{\tensor{1}}] \delta_{\sigma \sigma'},
\end{split}
\end{equation*}
where $\rho_i = \frac{1}{2 \sqrt{\lambda}} (1 - z^4_i)$. Then by restricting to the constraint surface, the last term disappears since $\C^{(0)} \approx 0$ and we end up with the following `weak' $r/s$-matrix algebra,
\begin{equation} \label{r-s alg H}
\{ L_{\text{H}\tensor{1}}, L_{\text{H}\tensor{2}} \} \approx [r^0_{\tensor{12}} - s^0_{\tensor{12}}, L_{\text{H}\tensor{1}}] \delta_{\sigma \sigma'} + [r^0_{\tensor{12}} + s^0_{\tensor{12}}, L_{\text{H}\tensor{2}}] \delta_{\sigma \sigma'} - 2 s^0_{\tensor{12}} \delta'_{\sigma \sigma'},
\end{equation}
where the matrices $r^0, s^0$ are related to $r,s$ as follows
\begin{equation} \label{r0s0}
r = r^0 + \mbox{\small $\frac{1}{2}$} (\rho_1 - \rho_2) C^{(00)}, \qquad s = s^0 + \mbox{\small $\frac{1}{2}$} (\rho_1 + \rho_2) C^{(00)}.
\end{equation}
This is precisely the relation between the $r,s$-matrices of \cite{SNM, Magro}. To see this, let us identify $r,s$ with the matrices $r^{\text{M}}, s^{\text{M}}$ in \cite{Magro} since we are using the same convensions as there. Let us also relate $r^0, s^0$ to the matrices $r^{\text{SNM}}, s^{\text{SNM}}$ in \cite{SNM} by setting $r^0 = - \frac{1}{4} r^{\text{SNM}}$ and $s^0 = \frac{1}{4} s^{\text{SNM}}$. The overall factor of $\frac{1}{4}$ and sign difference for the $r$-matrix have been discussed in \cite{Magro} and come down to conventions. With these identifications the relation \eqref{r0s0} then reads
\begin{subequations} \label{M vs SNM}
\begin{align}
-4 r^{\text{M}}_{\1\2} &= r^{\text{SNM}}_{\1\2} + (2 - z^4_1 - z^4_2) C^{(00)}_{\1\2},\\
4 s^{\text{M}}_{\1\2} &= s^{\text{SNM}}_{\1\2} + (z^4_1 - z^4_2) C^{(00)}_{\1\2},
\end{align}
\end{subequations}
in agreement with the results of \cite{SNM, Magro}.
\newpage

\section{Conclusions and Outlook}

By reconsidering the construction of the Lax connection for the Green-Schwarz superstring in $AdS_5 \times S^5$ from a purely Hamiltonian perspective we arrived at the Bena-Polchinski-Roiban connection extended by first and second class Hamiltonian constraints, exactly matching the one considered in \cite{Magro} where it was introduced in a rather \textit{ad hoc} way.

\paragraph{Extensions of BPR.} Our construction was broken down into two steps $(\text{I})$ \& $(\text{II})$: before demanding that our connection satisfies the zero curvature equation in the strong sense, we first imposed the requirement that it led only to first class integrals of motion which is very natural in the Hamiltonian setting. Both steps resulted in an extension of the (spatial component of the) Bena-Polchinski-Roiban connection:
\begin{itemize}
   \item[$(\text{I})$] Demanding that the spectral invariants of the monodromy all be first class required the extension of $L_{\text{BPR}}$ by the (partly) second class fermionic constraints $\C^{(1,3)}$ as
\begin{equation*}
L_{\text{BPR}} \rightsquigarrow L_{\text{H}} \equiv L_{\text{BPR}} + \frac{1}{2 \sqrt{\lambda}} (1 - z^4) \left( z^{-3} \C^{(1)} + z^{-1} \C^{(3)} \right).
\end{equation*}
Without this extension the Lax matrix $L_{\text{BPR}}$ gives rise to integrals whose action via the Poisson bracket do not preserve the constraint surface. The reason for this is that the constraint surface contains a second class part coming from $\C^{(1,3)}$. If we were to introduce a Dirac bracket which took care of these second class constraints then, after setting them strongly to zero, the action via the Dirac bracket of the integrals obtained from either $L_{\text{BPR}}$ or $L_{\text{H}}$ would both preserve the constraint surface. The Lax matrix $L_{\text{H}}$ can therefore be thought of as a Dirac extension of $L_{\text{BPR}}$ with respect to the second class parts of the fermionic constraints $\C^{(1,3)}$.
   \item[$(\text{II})$] Requiring the Lax connection to satisfy a zero curvature equation in the strong sense led to the further introduction of the first class bosonic constraint $\C^{(0)}$ as
\begin{equation*}
L_{\text{H}} \rightsquigarrow L_{\text{M}} \equiv L_{\text{H}} + \frac{1}{2 \sqrt{\lambda}} (1 - z^4) \C^{(0)}.
\end{equation*}
Without this second extension the Lax matrix $L_{\text{H}}$ is the spatial component of a weakly flat connection \---\ in other words flat only on the constraint surface. Fortunately, the weakly vanishing terms which appear when considering the flatness of $L_{\text{H}}$ could be accounted for by adding to $L_{\text{H}}$ an extra term in $\C^{(0)}$.
\end{itemize}

Notice that the first extension is of a fermionic nature and relates to $\kappa$-symmetry while the second extension has to do with the coset nature of $AdS_5 \times S^5$. This explains why these issues never arose in the context of bosonic strings formulated as principal chiral models \cite{b_attempts}.

\paragraph{Algebraic curve and finite-gap.} An interesting question one may ask is how the algebraic curve \cite{classification} is affected by the extensions. Since all the constraints vanish on the constraint surface by definition, they won't modify the algebraic curve associated to the Lagrangian equations of motion which only dictate the dynamics on the constraint surface. One may wonder however what the significance of the algebraic curve constructed from the strongly flat connection \eqref{strong zero curv} might be, and whether it has a dual gauge theory interpretation. In the same spirit, it would be interesting to apply the finite-gap construction to the strong zero curvature equation \eqref{strong zero curv} and determine whether the resulting action-angle variables are sensitive to the extensions.

\paragraph{Connection with pure spinors.} In \cite{Magro} the coefficients of the constraints $\C^{(1,3)}$ in \eqref{lax H} were fixed by comparison with the pure spinor Lax connection \cite{Vallilo}. Indeed, $L_{\text{H}}$ is exactly the spatial component of the pure spinor connection with the ghosts set to zero. In fact, introducing
\begin{equation*}
A^{(1)}_0 = \frac{1}{2} A^{(1)}_1 - \frac{1}{\sqrt{\lambda}} (\nabla_1 \Pi_1)^{(1)}, \qquad
A^{(3)}_0 = - \frac{1}{2} A^{(3)}_1 - \frac{1}{\sqrt{\lambda}} (\nabla_1 \Pi_1)^{(3)}.
\end{equation*}
which are the appropriate definitions of the temporal components $A^{(1,3)}_0$ in the Hamiltonian formalism of the pure spinor string \cite{Magro}, defining $A^{(0)}_0 = A^{(0)}_1$ as in \eqref{partial gauge} and noting that
\begin{equation*}
A^{(2)}_0 = - \frac{1}{\sqrt{\lambda}} (\nabla_1 \Pi_1)^{(2)}
\end{equation*}
from the constraint $\C^{(2)} = 0$ in the conformal gauge $\lambda^{\pm} = 1$, we can rewrite \eqref{L+-} as
\begin{align*}
L_+ &= A^{(0)}_+ + z A^{(1)}_+ + z^2 A^{(2)}_+ + z^3 A^{(3)}_+ + \rho \, \C^{(0)},\\
L_- &= A^{(0)}_- + z^{-3} A^{(1)}_- + z^{-2} A^{(2)}_- + z^{-1} A^{(3)}_-,
\end{align*}
where $A^{(i)}_{\pm} \equiv \frac{1}{2} (A^{(i)}_0 \pm A^{(i)}_1)$. Now apart from the term in $\C^{(0)}$ and the absence of ghosts, these are exactly the holomorphic and anti-holomorphic parts of the pure spinor Lax connection \cite{Vallilo}. It would therefore be of interest to construct the Lax connection for the pure spinor superstring within the Hamiltonian formalism, in the spirit of the present article, and compare the result to the Hamiltonian Lax connection \eqref{L+-} of the Green-Schwarz superstring obtained here.

\section*{Acknowledgments}

I am very grateful to M. Magro for many fruitful discussions motivating the work presented here and also for carefully reviewing the drafts of this paper. I also thank N. Beisert, N. Dorey and F. Spill for commenting on earlier drafts. This work was supported by the ANR grant INT-AdS/CFT (ANR36ADSCSTZ).

\end{document}